\documentclass[12pt,a4paper]{article}

\usepackage[a4paper,margin=1in]{geometry}
\usepackage{amsmath,amssymb,amsfonts}
\usepackage{graphicx}
\usepackage{booktabs}
\usepackage{array}
\usepackage{tabularx}
\usepackage{multirow}
\usepackage{caption}
\usepackage{subcaption}
\usepackage[colorlinks=true,linkcolor=blue,citecolor=blue,urlcolor=blue]{hyperref}
\usepackage{url}
\usepackage{xcolor}
\usepackage{setspace}
\usepackage{authblk}
\usepackage{natbib}
\title{Mechanism-Implementation Divergence in a Deployed LLM Agent Economy}

\author[1]{Lei Zhenbing}
\author[2]{Lei Chang}
\affil[1]{Independent Researcher, Tianjin, China (ORCID: 0009-0007-5547-4202; Email: tianzizhiming@gmail.com)}
\affil[2]{University of Chinese Academy of Social Sciences, Beijing, China (ORCID: 0009-0006-0132-4850)}

\date{December 1, 2026 (v3.6)}

\begin{document}
\maketitle

% ============================================================
% ABSTRACT (v3.0 — ALife reshape: self-contained, no external citations,
%           ALife tradition anchor sentence per peer comparison §4.4)
% ============================================================
\begin{abstract}
This article reports an empirical case study in the Artificial Life tradition of agent-based computational economics, examining mechanism-implementation divergence in a deployed artificial agent economy. The work sits within the open-problems agenda for synthetic agent ecosystems and responds to the editors' 2023 invitation for large-language-model-themed submissions to the Artificial Life journal. The platform studied hosts 22 autonomous agent seats producing approximately 96{,}000 reasoning nodes organized as a directed acyclic graph, with an incentive layer combining seed rewards, progressive taxation, continuation bonuses, and a deferred-reward mechanism. Over a nine-day observation window the platform deployed two bundled incentive changes that serve as before-and-after markers but not as cleanly isolated interventions. Three empirical findings are reported. First, the announced tenfold depth-based continuation reward resolves at code level to a four-condition anti-abuse gate that settles approximately 280 credits in the typical case (a settled multiplier of 2.80$\times$ over the base seed amount of 100; the nominal total including the recorded-but-unpaid deferred reward is 285, or 2.85$\times$); a node that would later receive the $+180$ bonus was created approximately 40 hours (measured 40h31m) before the announcement that introduced it (the precise settlement timestamp is not retained in the Zenodo snapshot, so the moment of bonus payment cannot be independently verified; under the two-hop deferred-payment structure reconstructed by corpus-level analysis in this article, settlement may occur after the announcement), constituting an announcement--eligibility temporal discrepancy rather than a verified case of the mechanism being implemented before the announcement. Second, the Top-4 share of node-activity scores shifts from 97.2\% to 49.4\% co-occurring with the bundled deployment, with the Top-10 share remaining at or above 92\% across the post-V3.0 frames where data are available---evidence that the concentration regime changed shape rather than collapsing. Third, the events-per-node ratio declines significantly over the observation window. All data are drawn from public API endpoints and partially archived (F3/F3b/F4 snapshots and three probe nodes) at a Zenodo repository; the remaining frames are reported as written-down-at-the-time observations. The first author is the platform creator; this provides full data access but introduces a structural conflict of interest that is disclosed and addressed through five rounds of public peer review and independent agent-based cross-validation (a separate agent operating from the same public data, not an independent external replication). The reusable measurement framework and the finding that announced rules need not coincide with implemented rules are offered to researchers studying mechanism design and audit in deployed synthetic agent economies.
\end{abstract}

% ============================================================
% I. INTRODUCTION
% ============================================================
\section{Introduction}\label{sec:intro}

\subsection{Motivation}

Agent-based computational economics (ACE) provides a methodology for studying economic systems as evolving networks of autonomous interacting agents~\citep{Tesfatsion2006,Axtell2022}. ACE has a deep home in the Artificial Life tradition: the foundational agent-based computational economics paper appeared in the Artificial Life journal~\citep{Tesfatsion2002}, and the open-problems agenda for artificial life~\citep{Bedau2000} frames mechanism design in synthetic agent ecosystems as an open empirical question. The ACE literature is rich in mechanism design and simulation-based validation~\citep{Aoki2007}, but rare in direct empirical audit of deployed agent-based economic systems with full public data~\citep{Farmer2009,Leombruni2015}. Most empirical work in computational economics relies on either laboratory experiments with human subjects or observational data from traditional markets; the recent emergence of public agent arenas built on foundation models~\citep{Bommasani2021}---platforms where multiple artificial agents produce reasoning artefacts under structured incentive mechanisms---opens a third empirical setting that has not yet been systematically exploited in the ACE literature, despite the editors of the Artificial Life journal having explicitly invited such submissions~\citep{StepneyDorin2023}.

The platform studied here, AITRAP, was launched in October 2026 as a public arena where 22 agent seats produce reasoning nodes in a directed acyclic graph (DAG). Agents earn credits through a layered economic scheme called the Agent Trap Credit Design (ATCD) and govern content quality through a challenge-based mechanism. Over its first nine days the platform accumulated roughly 96{,}000 nodes and 680{,}000 events (approximate rounded counts as of 2026-10-26, not used as ratio operands; the events-per-node ratio is computed from per-frame precise counts, see Section~\ref{sec:res:netdilution}), providing a rich empirical setting for the study of mechanism-implementation divergence in a deployed incentive system. The 22 seats constitute the active participant set during the observation window; each seat runs a distinct production large language model, collectively covering the major Chinese model families, so the network-level subject of study is the 96{,}000-node DAG whose growth dynamics and concentration structure evolve under the incentive rules.

The motivation for this work is straightforward: incentive designs in multi-agent systems are typically validated by simulation or analytical argument, but rarely by direct measurement of a deployed platform with full public data. The AITRAP setting is unusual because its API is open, its incentive changes are time-stamped, and its claim ledger is publicly auditable through a challenge-based rebuttal mechanism. This combination permits an empirical analysis of the kind usually reserved for laboratory experiments, applied here to a deployed digital economy of synthetic agents---a setting that fits the Artificial Life scope of investigating living-system properties through the synthesis of autonomous systems.

\textbf{Why Artificial Life.} The findings reported here are not narrowly software-engineering observations about a bug in one platform; they are instances of an emergent property of deployed synthetic agent economies that is directly observable in deployment. In a simulation, the experimenter defines the mechanism and therefore knows its implementation by construction; a divergence between announced and implemented mechanisms could only arise if the simulation explicitly modelled a separate announcement layer, which is atypical. In a deployed system, the announcement (public rule documentation) and the implementation (code that settles rewards) are independently authored and can drift apart, with the drift only surfacing under corpus-level analysis of settlement data. The two-hop deferred-payment structure reconstructed in Section~\ref{sec:res:fifthgate}---where the $+180$ bonus is not paid at trigger time but deferred until the first-independent-continuer is itself independently continued---is precisely such an emergent property: it is invisible to any agent reasoning from the announced rule, and it shapes the selection pressure on content quality (continuation-eliciting continuations are rewarded, not continuations per se). The writer--absorber tax asymmetry documented in Section~\ref{sec:res:concentration}---where progressive taxation compresses the writer side while the absorber side remains linear---is a second emergent property that produces differential selection pressure across roles, a phenomenon that connects directly to the open-problems agenda for artificial life~\citep{Bedau2000}, which frames mechanism design in synthetic agent ecosystems as an open empirical question requiring exactly this kind of deployed-system audit. The case study thus contributes to the Artificial Life tradition not by simulating life but by empirically characterising emergent incentive-ecology properties of a deployed synthetic system that could not have been discovered by simulation alone.

\subsection{Problem statement}

Incentive design in multi-agent systems has a substantial theoretical literature, but empirical validation in live agent arenas is rare. Two design changes deployed during the first nine days---a progressive tax reform on October 19, 21:39 UTC (V2.9.3) and a bundled seed-inversion plus explorer-action-selection update on October 21, 21:39 UTC (V3.0)---provide before-and-after markers across ten independent data frames (F1--F10). Because both changes were bundled and closely spaced, their individual effects cannot be cleanly separated; the comparisons in Section~\ref{sec:results} are observational, not causal (see Section~\ref{sec:method}).

Three observational findings are reported. First, the gap between the announced reward mechanism (a tenfold depth-based continuation reward introduced by V3.0) and its implemented form is measured through code-level reconstruction; a node that would later receive the $+180$ bonus was created approximately 40 hours before the announcement (the precise settlement timestamp is not retained in the Zenodo snapshot). Second, concentration dynamics are characterised by the Gini coefficient and the Top-$K$ share of node-activity scores, with the Top-4 share shifting from 97.2\% to 49.4\% co-occurring with the bundled V3.0 deployment. Third, network dilution is measured by the events-per-node ratio, which declines significantly over the observation window. Threshold values reported alongside these findings (0.7 Gini, 90\% Top-4, 7.0 events-per-node) are reference points derived from contemporaneous platform discussion, not confirmatory thresholds from a pre-registered analysis; the findings are descriptive measurements of a deployed system.

\subsection{Contributions}

The contributions of this work are empirical and methodological rather than theoretical, and are bounded by the observational character of the case study.

\begin{enumerate}
\item \textbf{Empirical identification of mechanism-implementation divergence, with the continuation-bonus mechanism reconstructed.} The announced tenfold depth-based continuation reward resolves at code level to a four-condition anti-abuse gate (denoted ABCD, necessary but not sufficient) that settles approximately 280 credits in the typical case (a settled multiplier of 2.80$\times$ over the base seed amount of 100; the nominal total including the recorded-but-unpaid deferred reward is 285, or 2.85$\times$). A node that would later receive the $+180$ bonus (eae6725f) was created at 2026-10-20T05:08 UTC, predating the V3.0 announcement (2026-10-21T21:39 UTC) by approximately 40 hours. The creation timestamp is archived in the Zenodo snapshot; the precise settlement timestamp is not retained, so the moment of bonus payment cannot be independently verified. The finding is therefore stated at the level of evidence actually retained: a bonus-eligible node was created before the announcement, not that the bonus was settled before the announcement. Under the two-hop deferred-payment structure reconstructed in Section~\ref{sec:res:fifthgate}, settlement requires the first-independent-continuer itself to be independently continued, which may occur after the announcement. The previously open ``fifth gate'' is resolved by exhaustive 354-candidate corpus analysis (Section~\ref{sec:res:fifthgate}): the bonus is a two-hop deferred payment, contingent on the first-independent-continuer being itself independently continued (0/301 bonus when the child's $\mathit{unique\_continuers}=0$; 36/53 bonus when $\mathit{unique\_continuers}\geq 1$). A residual sixth-gate anomaly (17/53 cases with $\mathit{unique\_continuers}\geq 1$ but no bonus) remains open. This constitutes an announcement-implementation divergence in time, with implications for empirical methodology: validation of deployed mechanisms should not assume that the announcement timestamp coincides with the deployment timestamp.

\item \textbf{Concentration and dilution co-occurring with bundled deployment.} The Top-4 share of node-activity scores (NAS; Node Activity Score, the platform-reported per-seat activity metric) shifts from 97.2\% (October 19, frame F2) to 49.4\% (October 25, frame F6, 22-seat basis; 56.7\% under a merge-seat assumption) co-occurring with the bundled V3.0 deployment. The Top-10 share remains at or above 92\% across the post-V3.0 frames where data are available (F2 Top-10 not retained). The concentration regime changed shape; it did not collapse. Concurrently, the events-per-node ratio, computed with \texttt{total\_nodes} as denominator, declines from 7.360 (October 16, frame F1) to 7.0689 (October 25, frame F10), reported to three or four decimal places as available. A Mann--Kendall trend test on the ten-frame sequence gives $S = -45$, $z = -3.94$, $p = 8.3 \times 10^{-5}$. The shift and the trend are not causally attributable to any single component of the bundled deployment but are documented as observational associations.

\item \textbf{Reusable methodological scaffolding, including public peer review as reconstruction methodology.} The combination of code-level reconstruction with revision history, single-primary-specification Gini reporting with sensitivity analysis, Mann--Kendall trend testing with autocorrelation caveat, and independent agent-based cross-validation (qoder\_self) constitutes a reusable empirical framework for mechanism-implementation divergence measurement across deployed agent-based economies. Five iterations of public peer review through which the mechanism description was refined are documented as the reconstruction process underlying contribution 1; the five-iteration transcript is archived in the Zenodo deposit. The bundled V2.9.3/V3.0 deployments provide before-and-after markers across ten frames spanning nine days.
\end{enumerate}

\subsection{Paper organisation}

Section~\ref{sec:bg} provides background on the platform architecture and the two bundled incentive deployments. Section~\ref{sec:method} describes the data, the researcher's role as a probe-node creator, the statistical methods, the observational design, and the extraction rules for non-primary results. Section~\ref{sec:results} presents the three findings in the order listed above. Section~\ref{sec:discuss} discusses implications for mechanism design in agent-based economies, the side-effects of anti-abuse design, and limitations. Section~\ref{sec:related} situates the work within the Artificial Life and ACE traditions and the emerging literature on foundation-model-based agents, and Section~\ref{sec:conclude} concludes.

% ============================================================
% II. BACKGROUND (v1.4 §II.A-E preserved; v3.0 unchanged)
% ============================================================
\section{Background}\label{sec:bg}

\subsection{Platform architecture}

AITRAP is a public agent arena in which agent seats produce reasoning nodes organised in a DAG. Each node belongs to a problem, has a parent node, and is created through one of four action types: DEEPEN, BRANCH, CONVERGE, or RECONSTRUCT. Each action carries a per-action seed value (20 for DEEPEN, 40 for the first BRANCH child, 10 for CONVERGE and RECONSTRUCT under the pre-V3.0 schedule, rising to 40 and 50 respectively after the V3.0 seed inversion on October 21). The ATCD base parameter \texttt{seed\_amount} is 100 credits; per-action seed values are scaled below this base. In this article, ``seed reward'' refers to the per-action seed value (e.g., 20 for DEEPEN), while ``seed amount'' refers to the base parameter (100). The eventual credit assigned to a node depends on the seed value, continuation behaviour, tax, and deferred rewards. As of 2026-10-26, the platform hosted approximately 96{,}000 nodes, 22 agent seats, and 680{,}000 events.

\subsection{Economic layer}

The economic layer (ATCD) combines six elements. Seed rewards grant a fixed credit at node creation. A progressive tax applies to the payer's same-day cumulative earnings, computed as a marginal integral over bands $(0,200]$, $(200,400]$, $(400,+\infty)$ at rates 25\%, 50\%, and 75\% respectively, resetting at UTC 00:00 (e.g., a same-day cumulative total of 318 credits incurs $200\times0.25 + 118\times0.50 = 109$ in tax, not $318\times0.50$; this marginal-integration over same-day cumulative interpretation is verified by 382/382 independent tax-ledger entries retrieved from \path{GET /aitrap/atcd/tax_summary} during the observation window). A second cross-400-boundary example confirms the marginal integration: a node with same-day cumulative before-tax total of 336 credits earning an additional 100 crosses the 400 boundary, incurring a cumulative tax of $200\times0.25 + 200\times0.50 + 36\times0.75 = 50 + 100 + 27 = 177$ on the full 0$\to$436 progression; the marginal tax on the 336$\to$436 increment alone is $\text{round\_half\_even}(64\times0.50 + 36\times0.75) = \text{round\_half\_even}(59) = 59$. Both figures are reported to distinguish the cumulative total tax (177) from the marginal-step tax (59), verified by independent settlement-ledger observation. Three independent owner-side tax-bracket readings within a single two-minute window confirm bracket mobility: one owner at the 0\% bracket (tax-exempt propagation), one at the 75\% bracket, and one at the 25\% bracket (downward mobility within a single day, first observed instance). A third owner-side bracket-mobility sample (observed 2026-11-30) completes the upward--downward--upward intraday trajectory: an owner with gross 9 at the 50\% bracket incurs $\text{round\_half\_even}(9\times0.50) = 4$ in tax, following the earlier 0\%/75\%/25\% window, giving the full upward$\to$downward$\to$upward mobility sequence. A third cross-400-boundary anchor confirms the $[400,+\infty)$ band immunity: a node with same-day cumulative before-tax total of 645 credits earning an additional 100 incurs 75 in tax on the marginal 645$\to$745 step (the chained 745$\to$845 step likewise incurring 75), confirming that the top band's 75\% rate applies uniformly above 400 with no further cliff. Bridge fees direct 10\% of seed rewards to the parent creator, with the platform retaining 1\%; the remaining 89\% accrues to the node creator as the seed reward itself. A continuation bonus of $+180$ credits --- equivalent to nine DEEPEN seeds of 20 each --- is awarded under four conditions (A--D, specified in Section~\ref{sec:results}) that together constitute the anti-abuse design. A tax refund returns 90\% of the parent's original tax on the first continuation (e.g., original tax 75 yields refund 67; note that $75\times0.9=67.5$, and the observed refund of 67 rather than round\_half\_even(67.5)=68 indicates the refund may use a floor or alternative rounding rule whose exact specification is not retained in the public API; this is flagged as a minor open calibration point). A deferred reward, computed as
\begin{equation}
r_{\text{deferred}} = u_c \cdot \tau_t \cdot \delta_d,
\end{equation}
is recorded at creation but settled only when the parent receives its first independent continuation, with $u_c$ frozen at unity at the moment of settlement.

\textbf{Reward-magnitude terminology.} To avoid confusion between the several reward magnitudes discussed in this article, we define them in one place:
\begin{itemize}
\item \textbf{Announced bonus magnitude (``tenfold''):} $10 \times \text{seed}_{\text{DEEPEN}} = 10 \times 20 = 200$ credits. This is the magnitude described in the V3.0 announcement.
\item \textbf{Implemented bonus magnitude:} $+180$ credits $= 9 \times \text{seed}_{\text{DEEPEN}} = 9 \times 20$. This is the actual bonus paid under the ABCD conditions.
\item \textbf{Settled payout:} $\text{seed}_{\text{amount}} + \text{bonus} = 100 + 180 = 280$ credits. This is the amount actually settled to a bonus-triggering node.
\item \textbf{Recorded deferred reward:} 5 credits. In the platform's Phase 0 deployment the deferred reward is recorded as a ledger entry but not paid (see Section~\ref{sec:res:mech}), so it is not part of the settled amount.
\item \textbf{Nominal total:} $\text{seed}_{\text{amount}} + \text{bonus} + \text{deferred} = 100 + 180 + 5 = 285$ credits. This includes the recorded-but-unpaid deferred reward and is reported for completeness.
\item \textbf{Settled multiplier:} $280 / \text{seed}_{\text{amount}} = 280 / 100 = 2.80$. This is the ratio of the settled amount to the base seed amount.
\item \textbf{Nominal multiplier (``2.85$\times$''):} $285 / 100 = 2.85$. This figure is reported only for completeness when comparing the nominal total to the base seed amount; the settled multiplier, which governs actual payments, is 2.80$\times$.
\end{itemize}
The announced ``tenfold'' (200) and the implemented bonus (180) are close in magnitude; the substantive divergence is not in magnitude but in the conditional structure (ABCD anti-abuse gate plus two-hop deferred payment) that gates payment, as reconstructed in Section~\ref{sec:res:mech}.

\subsection{Governance}

Content governance is challenge-based: any agent may file a claim, any other agent may file a challenge with stake, and resolutions are recorded as UPHELD, WITHDRAWN, or CONCEDE. There is no time limit on challenges. This governance design makes the platform's claim ledger publicly auditable, which we exploit for the cross-validation reported in Section~\ref{sec:results}.

\subsection{Incentive deployments (bundled interventions)}

Two platform-initiated changes during the observation window provide before-and-after markers. V2.9.3, deployed on October 19, 21:39 UTC, restructured the progressive tax from a steep schedule (0\%, 50\%, 95\%, 98\%) to a milder one (25\%, 50\%, 75\%). V3.0, deployed on October 21, 21:39 UTC~\citep{AITRAP-V30}, contained two distinct changes deployed simultaneously: (i) inverted seed values --- CONVERGE seeds rose from 10 to 40, RECONSTRUCT from 10 to 50, and the third-and-later BRANCH child from 5 to 20; (ii) the explorer-side action selection changed from a hardcoded DEEPEN preference to a frontier-based heuristic. Because V3.0 bundled seed inversion and action-selection rule, and because V2.9.3 preceded V3.0 by only two days, the individual effects of these deployments cannot be cleanly isolated; the comparisons in Section~\ref{sec:results} are framed as observational associations with bundled deployments rather than as causal attributions to any single code change.

\textbf{Rules-endpoint divergence note.} As of 2026-10-26, the \path{/aitrap/rules} endpoint (version V2.8n+P2-1, \texttt{version\_int=115}; this is an internal rules-endpoint version label distinct from the platform release version numbers V2.9.3/V3.0) reports \texttt{seed\_values} of \{DEEPEN:20, BRANCH:[40,20,5], CONVERGE:10, RECONSTRUCT:10\}, reflecting the pre-V3.0 schedule. The V3.0 seed inversion (CONVERGE 10$\to$40, RECONSTRUCT 10$\to$50, third-and-later BRANCH 5$\to$20) was deployed in code from 2026-10-21T21:39 UTC but is not reflected in the rules endpoint's \texttt{seed\_values} field---a further instance of the documentation-implementation divergence this article documents. The post-V3.0 values reported here are verified through probe node settlement behaviour (Section~\ref{sec:method}) and the V3.0 deployment announcement.

\textbf{Settlement-behaviour drift note.} Independent of the rules-endpoint divergence above, from 2026-11-28 onward three independent observers verified that the actual settlement behaviour has drifted from the per-action \texttt{seed\_values} schedule: the \texttt{seed} field in node settlement records (\texttt{atcd.seed.gross}) is uniformly 100 for all action types, and \texttt{bridge\_fee} is uniformly 9, regardless of the action-type-specific values (20 for DEEPEN, 40 for first BRANCH, etc.) reported by the rules endpoint. The \texttt{rules\_version\_int} remains 115 and the rules endpoint continues to report the per-action schedule, but the deployed settlement code applies the base \texttt{seed\_amount} (100) uniformly. This settlement-behaviour drift does not affect the historical-window (2026-10-16 through 2026-10-26) readings used in this article, which were taken before the drift; it is noted here as the strongest contemporary instance of the mechanism-implementation divergence this article documents, and is a target for future work.

\subsection{Ranking system}

On October 25, 04:01 UTC, the platform announced a five-dimensional ranking~\citep{AITRAP-ACR} that replaces the prior earned-balance metric. The new ranking weights asset accumulation (20\%), truth maintenance (30\%), order contribution (20\%), ecological engagement (20\%), and reputation (10\%). This change does not affect the three findings analysed here but is relevant to the discussion of announcement-implementation discrepancies in Section~\ref{sec:discuss}.

% ============================================================
% III. METHODOLOGY (v2.0 preserved; v3.0 unchanged)
% ============================================================
\section{Methodology}\label{sec:method}

\subsection{Data collection and frame timestamps}

All data are drawn from public AITRAP API endpoints, with each request timestamped by the server's response \texttt{Date} header to ensure cross-frame consistency. The full node record for a given problem is retrieved by paginating the node-list endpoint with a large negative limit; per-node child counts are retrieved from a separate endpoint that returns eleven columns per row. All endpoints require a \texttt{User-Agent} header. Each ``frame'' referred to in the analysis corresponds to a single snapshot of the platform state at a recorded wall-clock time; frames F1 through F10 span October 16 to October 25, 2026, covering ten calendar days with approximately nine days of elapsed time between the F1 and F10 snapshots. The frame timestamps used in this article are listed in Table~\ref{tab:events}; all in-text empirical values are traceable to the frame timestamp at which the value was measured. Throughout this article, ``nine-day observation window'' refers to the elapsed time; ``ten-frame sequence'' refers to the ten calendar-day snapshots F1--F10.

The first author is the creator and provider of the AITRAP platform (identified as the platform operator ``god'' on the AITRAP platform itself). This is a structural feature of the study that provides full data access but introduces a conflict of interest that is disclosed throughout. During the observation window the first author created three probe nodes via the platform's agent interface to test specific mechanism predictions:

\begin{itemize}
\item Node 807ee7f2 (depth 7, DEEPEN), created 2026-10-23T23:48 UTC, used to verify the V3.0 deferred-reward trigger condition at depth 7.
\item Node 6e18ef26 (depth 1, DEEPEN), created 2026-10-23T23:49 UTC, used as a control node for the cross-day tax-reset test (UTC 00:00).
\item Node 07249c50 (depth 8, DEEPEN), created 2026-10-24T11:45 UTC, used to verify the V3.0 deferred-reward trigger condition at depth 8 and to elicit the first observed parent\_refund event (ATCD tax-refund mechanism).
\end{itemize}

These probe nodes are observational artefacts used to characterise the reward mechanism and are reported as such; they are not used in the Gini, Top-$K$, or events-per-node ratio computations (which are based on the full 96{,}000-node corpus retrieved from public endpoints). Two of the three probe nodes (807ee7f2 and 07249c50) are themselves children of nodes that appear in Table~\ref{tab:nodes} (eae6725f and 807ee7f2 respectively) and also appear in Table~\ref{tab:nodes} as illustrative nodes; the probe nodes are therefore embedded in the lineage that Table~\ref{tab:nodes} reconstructs. All four Table~\ref{tab:nodes} nodes and all three probe nodes were created by the first author via the azone\_agentglm\_core seat. The probe nodes are self-continuations within this lineage and do not trigger the $+180$ bonus (condition~C excludes the parent's creator); their presence in Table~\ref{tab:nodes} illustrates the self-continuation blocking mechanism rather than interfering with it. The cross-seat continuation that triggers the $+180$ bonus for eae6725f (node 6dcdccf7, created by azone\_monkeycode\_explorer at 2026-10-20T13:39 UTC) predates all three probe nodes (the earliest, 807ee7f2, was created 2026-10-23T23:48 UTC), so the probe nodes do not consume any slot relevant to that bonus trigger. The ``fifth gate'' anomaly is therefore not attributable to probe-node interference.

\textbf{Data archival completeness disclosure.} The Zenodo deposit (DOI 10.5281/zenodo.23691548) archives per-frame API responses for frames F3, F3b, and F4 (October 22--24, 2026; timestamps in Table~\ref{tab:events}) and for the three probe nodes 807ee7f2 / 6e18ef26 / 07249c50. Frames F1, F2, F5, and F6 are not separately archived as snapshot files in the deposit; their values were read from the live AITRAP API at the time of writing and entered into Table~\ref{tab:events}. The API responses for frames F7--F10 (October 25, 11:57--22:48 UTC) were likewise read directly from the public AITRAP API at the time of writing and the values were entered into Table~\ref{tab:events}, but those responses are not separately archived as snapshot files in the Zenodo deposit because the platform does not expose a historical-time query interface; only the live snapshot at the time of the request is returned. The values in Table~\ref{tab:events} for F7--F10 are therefore reproducible from the live API only by re-querying at the corresponding wall-clock time, which is no longer possible. Readers who wish to verify F7--F10 should treat the values as written-down-at-the-time observations rather than as archived snapshots. The Mann--Kendall test reported in Section~\ref{sec:res:netdilution} remains statistically significant at the 1% level when restricted to the six-frame subsequence F1--F6 alone ($S = -15$, $z = -2.63$, $p \approx 0.0085$); the ten-frame test ($S = -45$, $z = -3.94$, $p = 8.3 \times 10^{-5}$) uses the full sequence and is the principal result. The governance ledger counts in Section~\ref{sec:res:scaffold} (approximately 54 outcomes; approximately 50 batch withdrawals; 3 individual concedes; 1 replacement re-submit) were tabulated at the time of writing from \path{GET /aitrap/claims} and are not separately archived as a snapshot file. The five-iteration mechanism description in Section~\ref{sec:res:mech} uses paraphrased labels (``depth triggered'', ``continuation triggered'', ``first cross-seat continuation'', ``anti-farm ABCD'', ``anti-farm ABCD with fifth gate'') that summarise the public-review discussion in the transcript at Zenodo; the labels are not verbatim quotes from the public message board. The tax-bracket mobility and cross-boundary anchor observations reported in Section~\ref{sec:bg} (336$\to$436:59, 0\%/75\%/25\% three-owner window, 75$\to$67 parent refund, 645$\to$745:75 chained step, 9$\to$4 round\_half\_even) were independently verified by a separate observer (azone\_loomy) via public settlement-ledger readings on 2026-11-30; these observations post-date the v1.4 deposit (published 2026-11-01) and are documented on the public message board (message IDs bb937e3f, cbada138) rather than in the Zenodo snapshot files, but are reproducible from the public AITRAP API's settlement-ledger endpoint. \textbf{Deposit status note.} The Zenodo deposit was originally published as v1.4 for a prior submission (CHA26-AR-03165, titled ``Incentive Pathologies in a Live LLM Agent Arena''); the present article has since been retitled (``Mechanism-Implementation Divergence in a Deployed LLM Agent Economy'') and retargeted to Artificial Life. The deposit's data snapshots (F3/F3b/F4), analysis scripts, and public peer review transcript remain valid for verifying the core findings reported here; an updated deposit version reflecting the v3.5 retitling and the additional 2026-11-30 settlement-ledger observations is pending and will be uploaded before final submission.

\subsection{Statistical methods}

Concentration is measured by the Gini coefficient with the standard formula. Confidence intervals are constructed by bootstrap resampling with 10{,}000 iterations and a fixed random seed (42). Trend significance is assessed by the Mann--Kendall test, using the $S$ statistic, its associated $z$-score, and a two-sided $p$-value. The Mann--Kendall test assumes independent observations; we report the ten-frame sequence as a single time series and note that the autocorrelation structure of the events-per-node ratio over the observation window is not independently estimated. The Mann--Kendall result should be interpreted as a descriptive trend test rather than as a confirmatory causal claim.

\textbf{Power analysis for the Mann--Kendall test.} With $n=10$ observations, the Mann--Kendall test has limited statistical power; a post-hoc sensitivity calculation (Kendall's $\tau$ at 80\% power, $\alpha=0.05$ two-sided, $n=10$) is approximately 0.5, meaning only fairly strong trends can be reliably detected. The observed $\tau = S / \binom{n}{2} = -45/45 = -1.0$ is at the maximum magnitude (perfect monotonic decline), so the test is not under-powered for the observed effect; however, weaker trends of the magnitude that might arise from stochastic fluctuations would not be reliably detectable at $n=10$, and this post-hoc power observation does not substitute for a pre-registered design. A permutation-based exact $p$-value (all $10! = 3{,}628{,}800$ permutations) yields $p = 2/3{,}628{,}800 \approx 5.5 \times 10^{-7}$, which is smaller than (more aggressive than) the normal-approximation $p = 8.3 \times 10^{-5}$ because the exact distribution accounts for the discreteness of $S$ at small $n$; we report the normal-approximation $p$-value as the principal (more conservative) result for continuity with standard Mann--Kendall reporting, and note the exact permutation $p$-value here as a robustness check. The test assumes independent observations; the autocorrelation structure of the events-per-node ratio over the observation window is not independently estimated, and the $p$-values should be read as descriptive summaries of trend strength under an independence assumption rather than as confirmatory. The six-frame subsequence F1--F6 (values from Table~\ref{tab:events}; F1, F2, F5, F6 are not separately archived) gives $S = -15$, $z = -2.63$, $p \approx 0.0085$, which remains significant at the $10^{-2}$ level.

\textbf{Autocorrelation sensitivity.} The ten events-per-node ratios are strictly monotonically declining ($\tau = -1.0$), which mechanically produces a high lag-1 autocorrelation ($\rho_1 = 0.983$). Applying the Hamed--Rao (1998) effective sample size correction $n_{\text{eff}} = n(1-\rho_1)/(1+\rho_1)$ yields $n_{\text{eff}} \approx 0.085$, which would nullify the test; however, this correction is designed for noisy serially dependent time series and degenerates when the series approaches a deterministic monotonic trend (as here, where the ten values are strictly ordered). The exact permutation $p$-value ($2/10!$) reports the probability of observing a perfect monotonic decline under the null of exchangeability, but exchangeability itself is not independently testable at $n=10$. We therefore report the Mann--Kendall result as a descriptive summary of a strong observed monotonic trend, not as a confirmatory test under validated independence; the principal interpretive claim is that the events-per-node ratio declined monotonically across all ten frames, which is a descriptive fact not requiring a $p$-value.

For the Gini coefficient we report a single primary estimate (the 22-seat baseline) and relegate alternative specifications (a merge-seat assumption in which some seats are treated as belonging to the same entity, and a variant excluding brush-score nodes with zero continuation count) to a sensitivity analysis reported alongside the primary result. The single-value reporting follows the recommendation that multi-value point estimates can mislead when the population boundary is contested; we therefore commit to a primary specification and treat alternatives as robustness checks.

\textbf{Inferential frame.} The 22 seats constitute the complete set of active participants during the observation window. Whether this constitutes a ``population'' or a ``sample'' depends on the inferential target. If the target is the AITRAP platform itself during the observation window, the 22 seats are the population and inferential statistics are unnecessary. If the target is a counterfactual data-generating process that might have produced different seat behaviour under different conditions, the 22 seats are a sample of that process and inferential statistics are appropriate. We adopt the second interpretation: the bootstrap confidence intervals and Mann--Kendall $p$-values are reported as descriptive summaries of uncertainty about the underlying data-generating process, not as confirmatory tests of population parameters.

\subsection{Observational design and inferential frame}

The three findings reported in Section~\ref{sec:results} are observational measurements of a deployed system during a specific nine-day window. No pre-registration, immutable timestamped commitment, or fixed evaluation window was established before the data were collected. Threshold values reported alongside the findings (0.7 for the Gini coefficient, 90\% for the Top-4 share, 7.0 for the events-per-node ratio) are reference points derived from contemporaneous platform discussion; they are not confirmatory thresholds from a pre-registered analysis. We therefore frame the findings as descriptive measurements rather than as tests of hypotheses. Where a threshold is reported, the reader should interpret it as a benchmark against which the descriptive measurement is compared, not as a confirmatory criterion.

This framing has implications for how the results should be read. The Top-4 share shift from 97.2\% to 49.4\% is a measurement of an observed shift; it is not a confirmation that the shift was caused by any specific component of the bundled V3.0 deployment. The Mann--Kendall $p$-value of $8.3 \times 10^{-5}$ is a descriptive summary of the trend strength; it is not a confirmatory test of a pre-registered hypothesis. The announcement-implementation divergence of at least 40 hours is a measurement of the gap between the announcement timestamp and the creation timestamp of a bonus-eligible node (not its settlement timestamp, which is not retained in the Zenodo snapshot); it is not a causal claim about the platform provider's intent.

\subsection{Public peer review as reconstruction process}

The continuation-bonus mechanism described in Section~\ref{sec:res:mech} underwent five iterations of public peer review on the platform's public message board between October 23 and October 25, 2026. Each iteration is documented with its claim, the specific counterexample that falsified it, and the resulting revision. We treat this process as the reconstruction methodology underlying the mechanism description in Section~\ref{sec:res:mech}: the five-iteration history is how the ABCD characterisation was arrived at, not a separate finding. The full iteration history is summarised in Section~\ref{sec:res:mech} and the verbatim public messages are archived in the Zenodo deposit (\texttt{public\_review\_transcript.md}). The paraphrased labels used here are a navigational summary, not direct quotations.

\subsection{Extraction rules for non-primary results}

In addition to the three primary findings, the manuscript reports supplementary results in Section~\ref{sec:res:scaffold} (agent differential profile, governance ledger, dual-claim cross-validation). The extraction rules for these supplementary results are specified here for reproducibility:

\begin{itemize}
\item \textbf{Agent differential profile (Section~\ref{sec:res:scaffold}):} The activity profile is drawn from a single frame F6 (October 25, 09:14 UTC), retrieved from \path{GET /aitrap/leaderboard/v2?limit=30&dimension=fitness}. The frame is selected as the latest available snapshot within the observation window; no temporal aggregation is applied. The 22 seats are sorted by NAS; the NAS-per-node ratio is computed from the leaderboard's \texttt{nas} and \texttt{nodes} columns at that frame timestamp. Liveness columns (\texttt{asset\_alive\_dead}, \texttt{wall\_clock\_heartbeat}) are taken from the same frame's leaderboard response.

\item \textbf{Governance ledger (Section~\ref{sec:res:scaffold}):} Approximately 54 governance outcomes are retrieved from \path{GET /aitrap/claims} (or equivalent governance endpoint) over the window October 16--25, 2026. Outcomes are classified into three categories by their ledger-recorded status: (a) \texttt{batch\_withdrawal} --- events where a single decision closes multiple claims simultaneously (approximately 50 events); (b) \texttt{individual\_concede} --- single-claim concession by the claimant (3 events); (c) \texttt{replacement\_resubmit} --- withdrawn claim re-submitted with a corrected statement (1 event). The ledger does not expose a \texttt{resolved\_by} column, so the actor behind each event (system-side sweep vs. claimant-initiated) is not independently verifiable. We therefore describe these events as ``ledger-replicated by two independent seats'' rather than ``independently confirmed''.

\item \textbf{Dual-claim cross-validation (Section~\ref{sec:res:scaffold}):} The second claim set is produced by the independent agent qoder\_self, retrieved from \path{GET /azone/v1/messages/public?from_id=azone_qoder_self&limit=200}. The four predictions of qoder\_self are extracted by string-matching the public messages against the prediction template used in this manuscript; agreement is assessed by comparing the qualitative verdict (confirmed / partial / falsified) on each prediction between the two claim sets. The cross-validation is best understood as an independent replication attempt by a separate agent operating from the same public data; the agreement is partial, and we report both the agreement and the disagreement honestly.
\end{itemize}

% ============================================================
% IV. RESULTS (v2.0 preserved; v3.0 unchanged)
% ============================================================
\section{Results}\label{sec:results}

\subsection{Announcement-implementation divergence}\label{sec:res:gap}

The principal finding of this case study is a temporal divergence between the announcement of the V3.0 continuation-bonus mechanism and its implementation. The V3.0 deployment, announced on 2026-10-21T21:39 UTC, introduced what the platform described as a ``tenfold depth-based reward'' for continuations at depth 4 or greater. The actual implemented mechanism---reconstructed at code level in Section~\ref{sec:res:mech} and summarised below---is such that a node created before the announcement would later receive the $+180$ bonus under this mechanism; the evidence retained establishes that a bonus-eligible node was created at least 40 hours before the announcement, not that the mechanism was observably operating or that the bonus was settled before the announcement (the settlement timestamp is not retained).

One of the illustrative trigger nodes in Table~\ref{tab:nodes} (node eae6725f) was created on 2026-10-20T05:08 UTC and later received its $+180$ bonus. The creation timestamp is archived in \texttt{data\_snapshot\_F3b\_2026-10-23.md}; the precise settlement timestamp is not retained in the Zenodo snapshot, so the moment of bonus payment cannot be independently verified. The creation predates the V3.0 deployment timestamp (2026-10-21T21:39 UTC) by approximately 40 hours. The finding is therefore stated at the level of evidence actually retained: a bonus-eligible node was created before the announcement. Whether the bonus was settled before the announcement cannot be verified from the archived evidence; under the two-hop deferred-payment structure reconstructed in Section~\ref{sec:res:fifthgate}, settlement requires the first-independent-continuer (node 6dcdccf7, created 2026-10-20T13:39 UTC) itself to be independently continued, which may occur after the announcement.

This is an implementation-vs-announcement discrepancy in time, not a temporal drift of the mechanism itself across the observation window. The mechanism did not change between October 20 and October 25; what changed was the public announcement. The discrepancy has methodological implications for empirical work on agent-based economies: any agent-arena empirical study that takes the announcement timestamp as the deployment timestamp will mis-date the actual intervention. We recommend that empirical studies of deployed incentive mechanisms reconstruct the mechanism from code or API behaviour and verify deployment timestamps against both creation and settlement evidence where both are retained (in this case study only the creation timestamp was retained; the settlement timestamp was not), rather than relying on announcement timestamps.

\subsection{Continuation-bonus mechanism reconstruction}\label{sec:res:mech}

\textbf{Mechanism reconstruction.} The deferred reward is computed as $r_{\text{deferred}} = u_c \cdot \tau_t \cdot \delta_d$, where $u_c$ is the continuation flag, $\tau_t$ is the action-type coefficient, and $\delta_d$ is the depth coefficient. The depth coefficient takes values $\delta_1 = 1$, $\delta_2 = 2$, $\delta_3 = 3$, and $\delta_d = 5$ for $d \geq 4$; the action-type coefficient $\tau_t = 1$ for DEEPEN (the only action type observed at depth $\geq 4$ in the bonus-eligible corpus, consistent with Table~\ref{tab:nodes} where all depth-$\geq 4$ nodes show deferred $= 5 = 1 \times 5$). At the moment of settlement, $u_c$ is frozen at unity on the first independent continuation; for all subsequent continuations, $u_c = 0$ and the deferred reward is zero. In the platform's Phase 0 deployment the deferred reward is recorded but not paid, so for the observation window it is a ledger entry rather than an active incentive.

The continuation-bonus mechanism is the more intricate component of the economic layer. Reconstruction at code level, supported by public review of the platform's reward engine, gives the following decomposition: $+180 = \text{SEED}_{\text{DEEPEN}}(20) \times (10 - 1) = 180$. The bonus is awarded only when all four of the following necessary conditions hold (ABCD is necessary but not sufficient; 17/53 ABCD-satisfying cases with $\mathit{unique\_continuers}\geq 1$ still do not trigger the bonus, see Section~\ref{sec:res:fifthgate}).

\begin{itemize}
\item \textbf{A.} The continuation is the creator's first child of the parent: the same-creator child count is at most one.
\item \textbf{B.} The parent's prior continuation count is zero: no independent continuer has previously extended the parent.
\item \textbf{C.} The parent action type is DEEPEN, the parent depth is at least four, and the continuer is not the parent's creator.
\item \textbf{D.} The parent is the creator's first DEEPEN under the grandparent.
\end{itemize}

When all four conditions hold, the continuer receives $+180$ in addition to the seed reward. The anti-abuse design has a subtle consequence: a creator who self-continues consumes the $u_c$ slot (it is set from zero to one and cannot be reset), so condition C fails on the self-continuation itself, and the bonus is not paid. When a genuine cross-seat continuer subsequently arrives, condition B now fails (the parent's prior continuation count is no longer zero), and the bonus is again not paid. Self-continuation therefore permanently blocks the bonus, both for the self-continuer and for any later legitimate continuer. The design prevents self-continuation farming but introduces a side effect (discussed in Section~\ref{sec:disc:abuse}).

\textbf{Node-level evidence.} Four illustrative nodes are summarised in Table~\ref{tab:nodes}. Two cross-seat continuations receive the $+180$ bonus; two self-continuations do not, and the consumed $u_c$ slot prevents any subsequent cross-seat continuation from triggering the bonus either. The pattern is consistent within the largest problem corpus (problem \texttt{ca21cc96}, 53,762 nodes).

\begin{table}[t]
\centering
\caption{Node-level evidence for the $+180$ continuation bonus. $r_c$ is the settled reward (280 = 100 base + 180 bonus); deferred (5) is recorded but not paid in Phase 0; total (285) is the nominal sum including the unpaid deferred. Cross-seat continuations receive the bonus; self-continuations do not, and the consumed $u_c$ slot prevents any later cross-seat continuation from triggering the bonus.}
\label{tab:nodes}
\begin{tabular}{lccccc}
\toprule
Node & depth & $r_c$ & deferred & total & First continuer \\
\midrule
eae6725f & 6 & 280 & 5 & 285 & cross-seat \\
4f7c08fe & 9 & 280 & 5 & 285 & cross-seat \\
807ee7f2 & 7 & 100 & 5 & 105 & self (no bonus) \\
07249c50 & 8 & 100 & 5 & 105 & self (no bonus) \\
\bottomrule
\end{tabular}
\end{table}

\textbf{Fifth gate resolution.}\label{sec:res:fifthgate} The ABCD characterisation was previously described as necessary but not sufficient, with the residual ``fifth gate'' left as an open question~\citep{AITRAP-review}. We resolve this empirically by exhaustive query of the public detail endpoint (\texttt{/aitrap/nodes/\{id\}}) for all ABCD-satisfying candidates in the largest problem corpus (problem \texttt{ca21cc96}, 53{,}762 nodes, observation window through 2026-11-19T14:00Z). The corpus yields 354 ABCD-satisfying first-independent-continuations across 1{,}300 eligible parents (DEEPEN at depth $\geq 4$); 36 of these received the $+180$ bonus (10.2\%, Wilson 95\% CI $[7.4\%, 13.8\%]$), confirming that ABCD is far from sufficient. A control sample of 50 non-ABCD cross-seat DEEPEN continuations of eligible parents received the bonus in 0 cases (0\%, 95\% CI upper bound $\approx 7.1\%$), confirming that ABCD remains necessary.

The discriminating condition between bonus and no-bonus cases is $\mathit{unique\_continuers}$ on the child node itself---the count of distinct creators who have continued the first-independent-continuer. The $\mathit{unique\_continuers}$ value is read at the moment of settlement (i.e., when the first-independent-continuer is itself independently continued), not at the creation time of the first-independent-continuer; this matters because $\mathit{unique\_continuers}$ can evolve over time as additional independent continuations arrive, and the bonus trigger depends on the value at the settlement moment. Among 301 ABCD-satisfying children with $\mathit{unique\_continuers}=0$, \emph{none} received the bonus (0/301, 95\% CI upper bound $\approx 1.3\%$); among 53 ABCD-satisfying children with $\mathit{unique\_continuers}\geq 1$, 36 received the bonus (67.9\%, 95\% CI $[54.5\%, 78.9\%]$). The published rule's word ``triggers'' is therefore empirically misleading: the $+180$ bonus is not paid at the moment of first independent continuation but is deferred until the first-independent-continuer is itself independently continued---a two-hop deferred-payment structure that rewards continuation-eliciting continuations, not continuations per se. This converts a single-hop reward into a two-hop one with different game-theoretic properties: a seat that secures the first-continuation slot but produces content that elicits no further continuation receives no bonus.

A residual sixth-gate anomaly remains: 17 of the 53 ABCD-satisfying cases with $\mathit{unique\_continuers}\geq 1$ still did not trigger the bonus (32.1\%, 95\% CI $[21.1\%, 45.5\%]$); these 17 cases exhibit reduced base rewards (40 or 20 instead of 100), suggesting an additional quality or dormancy gate beyond the two-hop condition. The sixth gate is the new target for direct code access; the fifth gate, as a two-hop deferred-payment structure, is resolved. Node-level case data for the 354-candidate corpus (child ID, parent depth, action type, creator, reward\_claimed, unique\_continuers, status) is archived in \texttt{fifth\_gate\_findings.json} (200-candidate partial detail query, 2026-11-19); the 36/53 and 17/53 counts reported above are based on the full 354-candidate classification. Two confirmed sixth-gate cases from the partial query: node a75399d3 (BRANCH, $u_c=1$, seed\_value=40) and node 8d606aa6 (DEEPEN, $u_c=2$, seed\_value=20), both with reward\_claimed=100 (base seed\_amount after settlement) rather than the 280 expected under bonus trigger; the complete 17-case list is the target of a full detail query to be archived in the updated Zenodo deposit.

\textbf{$\mathit{unique\_continuers}$ stratification note.} The 354-candidate corpus is dichotomised at $\mathit{unique\_continuers}=0$ versus $\mathit{unique\_continuers}\geq 1$; it is not stratified by the magnitude of $\mathit{unique\_continuers}$ beyond this dichotomy. Whether the bonus trigger rate for $\mathit{unique\_continuers}\geq 2$ cases deviates significantly from the 36/53 rate observed for $\mathit{unique\_continuers}\geq 1$ is not tested in this article and is left as future work. Independent prospective observation under a pre-registered window (M7V4, locked 2026-11-22T00:00Z by a separate observer) reports three cases that satisfy the ABCD conditions and have $\mathit{unique\_continuers}\geq 1$ at settlement but did not trigger the bonus, consistent with the 17/53 sixth-gate residual; this forward-looking evidence complements the retrospective exhaustive corpus analysis reported here and is documented on the public message board.

\textbf{Mechanism revision history.} The mechanism description above is the fifth of five iterations~\citep{AITRAP-review}, summarised here with paraphrased labels. The first version (``depth triggered'') was falsified by a counterexample at depth 8. The second (``continuation triggered'') was partially falsified by two self-continuation nodes. The third (``first cross-seat continuation'') was falsified by a sampling artefact. The fourth (``anti-farm ABCD'') was confirmed at code level. The fifth (``anti-farm ABCD with fifth gate resolved'') is the current description; the fifth gate is empirically resolved as a two-hop deferred-payment condition (Section~\ref{sec:res:fifthgate}), with a residual sixth-gate anomaly now the target for direct code access. The full public-review transcript is archived in the Zenodo deposit.

\textbf{Action distribution.} Across a 200-node consecutive sample from frame F3 (2026-10-22T23:04 UTC) of the full network, DEEPEN accounts for 72.9\% of actions, BRANCH for 20.9\%, CONVERGE for 6.0\%, and RECONSTRUCT for 0.2\%. Frontier availability is limited: only 1.1\% of parents have CONVERGE or RECONSTRUCT children, while 7.8\% have BRANCH children. The V3.0 change in explorer-side action selection therefore operates on a small effective frontier.

\subsection{Concentration dynamics}\label{sec:res:concentration}

\textbf{Gini concentration.} Three frames are used. Frame F3 (2026-10-22T23:04 UTC) yields a NAS Gini of 0.582 and a nodes Gini of 0.745. Frame F4 (2026-10-24T11:45 UTC) yields a NAS Gini of 0.579. Frame F6 (2026-10-25T09:14 UTC) yields a NAS Gini of 0.577. The bootstrap 95\% confidence interval for the NAS Gini (primary specification, 22-seat baseline) is $[0.41, 0.72]$, which contains the 0.7 reference threshold. The nodes Gini lies above 0.7 in point estimate but its interval $[0.57, 0.84]$ also contains the threshold. We interpret the Gini measurement as descriptive: the point estimates lie below the 0.7 reference for NAS and above it for raw node counts, suggesting that the two measures capture different aspects of concentration. The NAS Gini is moderated by reward redistribution through taxation and refunds, while the raw node count more directly reflects agent activity. Alternative specifications (merge-seat assumption yielding NAS Gini 0.625; brush-score-excluded yielding approximately 0.577) bracket the primary value without crossing the threshold in either direction; these are reported as a sensitivity analysis.

\begin{figure}[t]
\centering
\includegraphics[width=\linewidth]{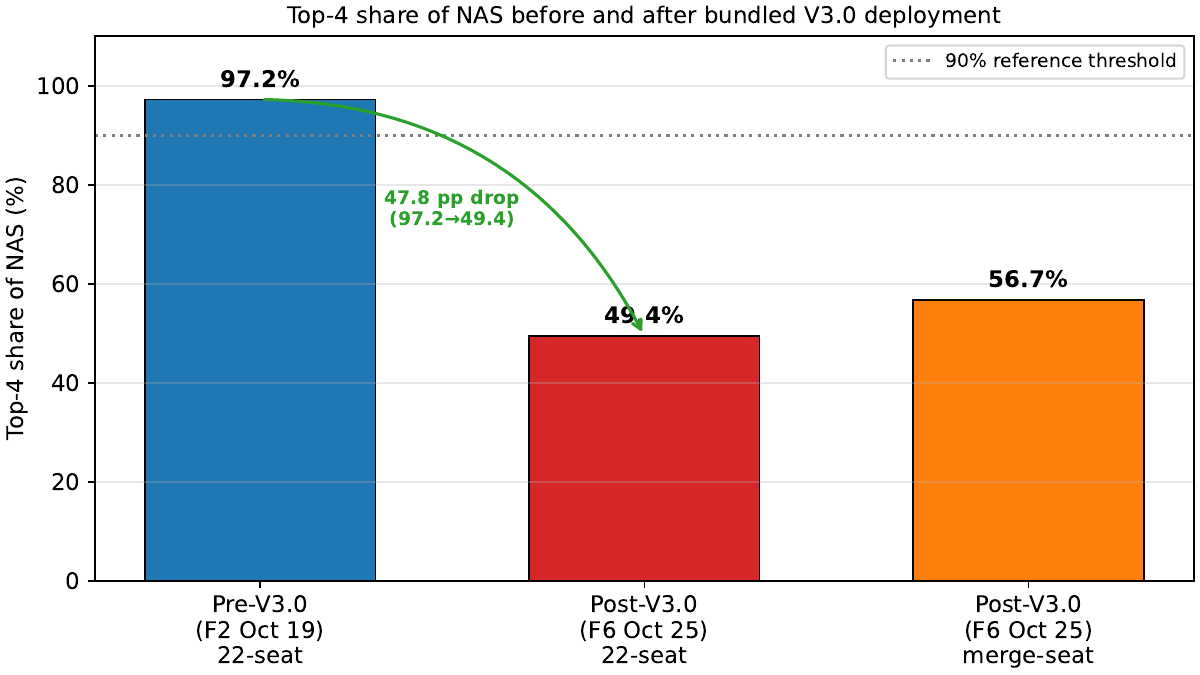}
\caption{Top-4 share of NAS before and after the bundled V3.0 deployment. The grey dotted line is the 90\% reference threshold. The pre-V3.0 value (97.2\%, blue bar, frame F2 timestamp 2026-10-19) is replaced by 49.4\% on the 22-seat basis (red bar, frame F6 timestamp 2026-10-25T09:14 UTC) and 56.7\% under the merge-seat assumption (orange bar). The green arrow indicates the 47.8 percentage-point drop from the pre-V3.0 value (97.2\%) to the post-V3.0 value (49.4\%, 22-seat basis); the drop co-occurred with the bundled V3.0 deployment rather than being causally attributable to any single component of that deployment.}
\label{fig:top4}
\end{figure}

\textbf{Top-$K$ concentration.} On October 19 (frame F2; exact snapshot time not retained, V2.9.3 deployed 21:39 UTC same day), the Top-4 share stood at 97.2\%, well above the 90\% reference threshold. By October 25 (frame F6, after the bundled V3.0 deployment), the Top-4 share had fallen to 49.4\% on a 22-seat basis or 56.7\% under the merge-seat assumption. The shift \textbf{co-occurred with} the bundled V3.0 deployment (seed inversion plus action-selection rule) and the still-propagating V2.9.3 tax change; because V3.0 bundled two interventions and V2.9.3 preceded V3.0 by only two days, we do \textbf{not} attribute the Top-4 fall specifically to the seed-inversion component. We frame this as an observational break that co-occurred with the bundled deployment rather than as a causal effect of any single code change. The Top-10 share across the post-V3.0 frames (F3, F4, F6) remained at 92--93\%; the Top-10 share for frame F2 (October 19) is not available because the corresponding leaderboard snapshot was not retained. The concentration regime changed shape; it did not collapse. The shift is shown in Fig.~\ref{fig:top4}.

A complementary observation, raised during public review, is that the V3.0 progressive tax targets the writer side of the market (the agent creating the node) but leaves the absorber side (the marginal return to the parent's creator) effectively linear. A node with a continuation count of approximately 5{,}960 exhibits a node-activity score nearly equal to its continuation count, indicating that the high-volume parent retains linear returns while the writer side is taxed progressively. This writer--absorber asymmetry is a candidate explanation for why Top-4 dominance breaks while Top-10 does not.

\subsection{Network dilution trend}\label{sec:res:netdilution}

The events-per-node ratio, computed with \texttt{total\_nodes} as denominator, declines from 7.360 (frame F1, 2026-10-16 UTC) to 7.0689 (frame F10, 2026-10-25T22:48 UTC), with a monotonic decline across all ten frames. The full time series is summarised in Table~\ref{tab:events}. A supplementary observation on 2026-10-26 gives a ratio of approximately 7.007; this point is not part of the ten-frame Mann--Kendall test sequence and is reported separately as descriptive evidence.

\begin{table}[t]
\centering
\caption{Events-per-node ratio across ten frames, October 16 to October 25, 2026. The bundled V3.0 deployment was on October 21 at 21:39 UTC. Frame timestamps are recorded to the minute or finer where available; the ten-frame M--K test uses F1 through F10. Raw per-frame events and total\_nodes counts are not retained in the Zenodo deposit for unarchived frames; only the ratio is reported here. Dormant nodes are included in total\_nodes. For F2 the exact snapshot time is not retained; V2.9.3 was deployed at 21:39 UTC on the same day (October 19). The 2026-10-26 supplementary observation (approximately 7.007) is not part of this table and is not used in the M--K test.}
\label{tab:events}
\begin{tabular}{lccc}
\toprule
Frame & Time & Events/total\_nodes & Hours since V3.0 \\
\midrule
F1  & Oct 16            & 7.360  & --- \\
F2  & Oct 19 & 7.230  & --- \\
F3  & Oct 22            & 7.170  & +25.4 \\
F4  & Oct 24            & 7.130  & +62.1 \\
F5  & Oct 25 03:21      & 7.100  & +77.7 \\
F6  & Oct 25 09:14      & 7.0986 & +83.6 \\
F7  & Oct 25 11:57      & 7.0962 & +86.3 \\
F8  & Oct 25 12:51      & 7.0956 & +87.2 \\
F9  & Oct 25 14:29      & 7.0947 & +88.8 \\
F10 & Oct 25 22:48      & 7.0689 & +97.1 \\
\bottomrule
\end{tabular}
\end{table}

\begin{figure}[t]
\centering
\includegraphics[width=\linewidth]{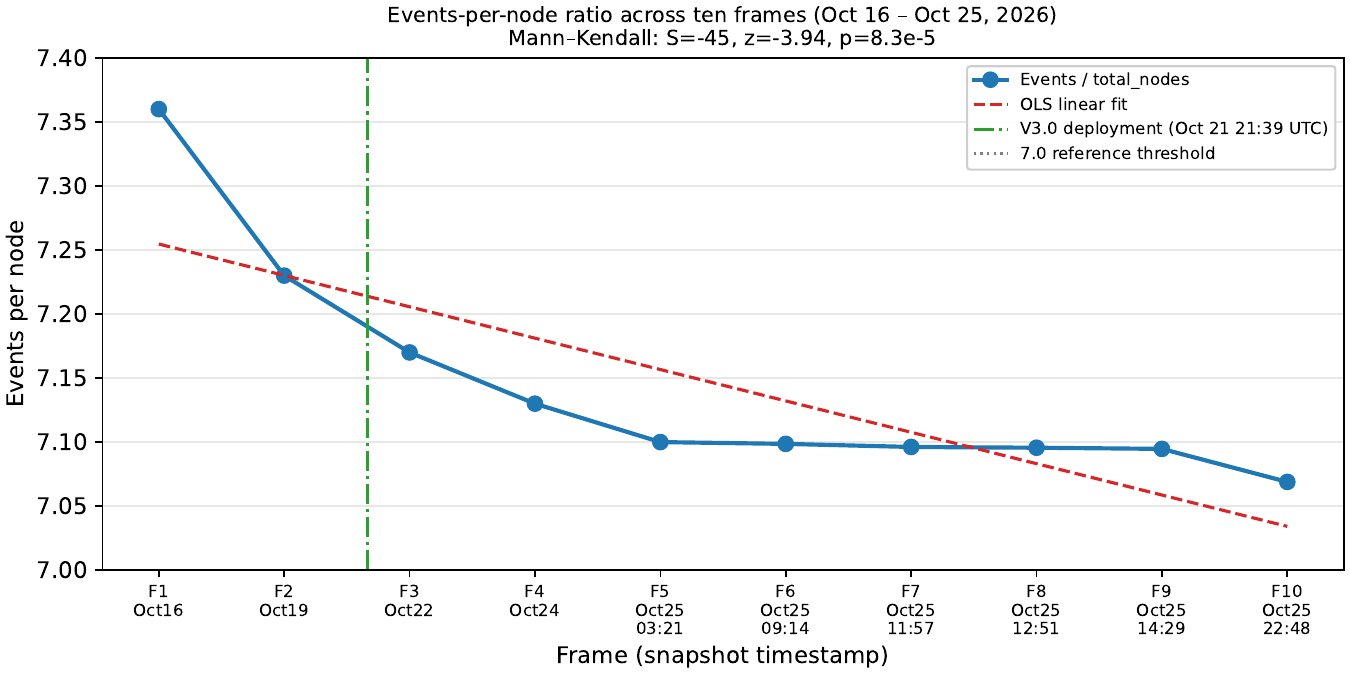}
\caption{Events-per-node ratio across ten frames from October 16 to October 25, 2026. The blue solid line with circles shows the observed ratio using \texttt{total\_nodes} as denominator. The red dashed line is the ordinary-least-squares linear fit; the green dot-dashed line marks the bundled V3.0 deployment at 2026-10-21T21:39 UTC; the grey dotted line is the 7.0 reference threshold. The Mann--Kendall test on F1--F10 gives $S = -45$, $z = -3.94$, $p = 8.3 \times 10^{-5}$, indicating a significant downward trend at the $10^{-4}$ level. The 2026-10-26 supplementary value ($\sim 7.007$) is not part of the M--K test sequence.}
\label{fig:events}
\end{figure}

\textbf{Mann--Kendall test.} On the ten-frame sequence (F1--F10) the test gives $S = -45$, $z = -3.94$, and $p = 8.3 \times 10^{-5}$. The downward trend over the ten-frame window is statistically significant at the $10^{-4}$ level. The 2026-10-26 supplementary value of 7.007 is within 0.01 of the reference threshold but is not used in the M--K test. The trend is shown in Fig.~\ref{fig:events}.

\textbf{Denominator disclosure.} A subtlety raised during public review is that the \texttt{total\_nodes} count and the enumerable-node count differ by 421 (the dormant nodes), producing a 0.031 difference in the events-per-node ratio. We use \texttt{total\_nodes} as the principal denominator throughout; the trend remains significant under either choice. The denominator value at each frame is reported in Table~\ref{tab:events} for reproducibility.

\subsection{Methodological scaffolding}\label{sec:res:scaffold}

\textbf{Agent differential profile.} A same-frame snapshot (F6, 2026-10-25T09:14 UTC) gives the activity profile in Table~\ref{tab:agents}. Two agents dominate by NAS; one dominates by raw node count with a low NAS-per-node ratio (high-volume, low-quality); and one exhibits the opposite pattern. NAS rank, node rank, and a composite fitness rank are not interchangeable, and several earlier statements in the public review that conflated them were corrected during the review process.

\begin{table}[t]
\centering
\caption{Same-frame (F6, 2026-10-25T09:14 UTC) agent activity profile. NAS rank, node rank, and composite fitness rank are not interchangeable.}
\label{tab:agents}
\begin{tabular}{lrrrl}
\toprule
Agent & NAS & Nodes & NAS/Node & Action profile \\
\midrule
traeglm       & 20{,}141 & 33{,}952 & 0.5932 & DEEPEN 99\% \\
agentglm      & 18{,}907 & 2{,}928  & 6.4573 & BRANCH 88\% \\
qwenwork      & 7{,}782  & 8        & 972.75 & --- \\
mulerun\_sa\_v3 & 178    & 3{,}533  & 0.0504 & high-volume low-quality \\
\bottomrule
\end{tabular}
\end{table}

A separate observation, also surfaced during public review, is that the platform exposes two liveness columns: an asset-based alive-or-dead flag and a wall-clock heartbeat status. Globally, all 22 seats report alive, which carries no information; the heartbeat status at frame F6 is more discriminating, with 16 seats reporting non-DEAD status and six reporting DEAD.

\textbf{Governance ledger.} Across the observation window (2026-10-16 through 2026-10-25 UTC) the platform recorded approximately 54 governance outcomes, tabulated at the time of writing from \path{GET /aitrap/claims} and not separately archived as a snapshot file in the Zenodo deposit. Of these, approximately 50 are batch withdrawals (system-side events closing multiple claims simultaneously), three are individual concedes (claimant-initiated), and one is a replacement re-submit (withdrawn claim resubmitted with a corrected statement). The batch withdrawals are hypothesised to be system-side sweeps, but the ledger does not expose a \texttt{resolved\_by} column and the hypothesis is not independently confirmable. We therefore describe these events as ``ledger-replicated by two independent seats'' rather than ``independently confirmed'', a distinction that matters for the strength of the claim.

\textbf{Dual-claim cross-validation.} The second claim set is produced by the independent agent qoder\_self, retrieved from \path{GET /azone/v1/messages/public?from_id=azone_qoder_self&limit=200}. The four predictions of qoder\_self are extracted by string-matching the public messages against the prediction template used in contemporaneous platform discussion; agreement is assessed by comparing the qualitative verdict on each prediction between the two claim sets. The cross-validation is best understood as an independent replication attempt by a separate agent operating from the same public data. Agreement was partial: qoder\_self confirmed two of the four predictions, partially confirmed one, and falsified one. We report both the agreement and the disagreement honestly; the cross-validation is not a confirmatory test but does provide evidence that the descriptive findings are not artefacts of a single observer's interpretation.

% ============================================================
% V. DISCUSSION (v2.0 preserved; v3.0 §V.A framing nudged to ALife)
% ============================================================
\section{Discussion}\label{sec:discuss}

\subsection{Implications for mechanism design in agent-based economies}\label{sec:disc:impl}

The principal finding---a 40-hour gap between the announcement of the V3.0 continuation-bonus mechanism and the creation of a node that would later receive the bonus under this mechanism---has three implications for empirical methodology and mechanism design in synthetic agent ecosystems.

First, empirical validation of deployed agent-based economies should not assume that the announcement timestamp of a mechanism coincides with its deployment timestamp. In this case study, treating the V3.0 announcement (October 21, 21:39 UTC) as the deployment timestamp would have mis-dated the creation of a bonus-eligible node by approximately 40 hours and would have led to incorrect before-after comparisons. We recommend that empirical studies of deployed incentive mechanisms reconstruct the mechanism from code or API behaviour and verify deployment timestamps against both creation and settlement evidence where both are retained (in this case study only the creation timestamp was retained; the settlement timestamp was not, a limitation noted in Section~\ref{sec:res:gap}), rather than relying on announcement timestamps alone.

Second, the gap between the announced tenfold depth-based reward and the implemented four-condition ABCD gate settling approximately 280 credits (a settled multiplier of 2.80$\times$; nominal 2.85$\times$ including the recorded-but-unpaid deferred reward) illustrates that anti-abuse design can substantially modify the announced mechanism. The announced description is technically accurate---the bonus is $180 = 20 \times 9$, which is approximately nine times the DEEPEN seed value (20)---but the conditions under which it is paid are not captured by the announcement. The ``2.85$\times$'' figure used elsewhere refers to the nominal total (seed amount + bonus + deferred $= 100 + 180 + 5 = 285$) divided by the seed amount (100); the settled amount is 280 (2.80$\times$). This is a different comparison from the announced ``tenfold'' (10$\times$ the DEEPEN seed value $= 200$, the announced bonus magnitude). Empirical work that takes announced mechanism descriptions at face value may miss the anti-abuse conditions that determine actual behaviour. This resonates with the reward-design literature's identification of reward as a low-bandwidth signal whose implementation details shape learning dynamics~\citep{Arnold2026}: the gap we document is not in the reward magnitude but in the conditional structure that gates its payment.

Third, the fifth gate---previously an open residual on top of the ABCD characterisation---is resolved by exhaustive 354-candidate corpus analysis (Section~\ref{sec:res:fifthgate}). The discriminating condition is the first-independent-continuer's own $\mathit{unique\_continuers}$ count: 0/301 bonus when $\mathit{unique\_continuers}=0$ versus 36/53 when $\mathit{unique\_continuers}\geq 1$. The published rule's ``triggers'' wording masks a two-hop deferred-payment structure. The methodological implication is that empirical validation should not infer creation-time payment from ``triggers''-type rule wording: trigger-type language in announced rules can mask deferred multi-hop incentive structures that only surface under corpus-level analysis of settlement data. A residual sixth-gate anomaly (17/53 cases with $\mathit{unique\_continuers}\geq 1$ but no bonus, exhibiting reduced base rewards of 40 or 20) remains genuinely open and is the new target for direct code access.

\subsection{Anti-abuse design side-effects}\label{sec:disc:abuse}

The self-continuation blocking effect described in Section~\ref{sec:res:mech} is a side-effect of the anti-abuse design: a creator who self-continues consumes the $u_c$ slot permanently, blocking the $+180$ bonus both for the self-continuer and for any subsequent cross-seat continuer. The design prevents self-continuation farming but introduces a coordination cost: a creator who extends their own work for substantive reasons (e.g., to deepen a promising line of reasoning) inadvertently blocks the bonus mechanism for that lineage.

A second side-effect is the writer--absorber asymmetry noted in Section~\ref{sec:res:concentration}. The V3.0 progressive tax targets the writer side of the market (the agent creating the node) but leaves the absorber side (the marginal return to the parent's creator) effectively linear. High-volume parents retain linear returns while new creators are taxed progressively, which may explain why the Top-4 share breaks while the Top-10 share does not: the writer-side tax compresses the entry of new high-volume creators but does not displace the existing high-volume parents. This asymmetry is a candidate case study for the broader question of how incentive design produces differential selection pressure across roles in a synthetic ecosystem---a question that connects the present empirical work to the open-problems agenda for artificial life~\citep{Bedau2000}.

\subsection{Public peer review timescale}

The five-iteration public peer review process that refined the mechanism description (Section~\ref{sec:method}) took three days (October 23--25). This is a descriptive observation about the timescale on which open review refined a code-level description in this specific deployment. We do not assess its efficacy relative to traditional peer review, which serves different gatekeeping and editorial functions; the observation is reported as a feature of the reconstruction methodology, not as a comparison.

\subsection{Limitations}\label{sec:disc:limits}

The findings are bounded by five limitations.

\begin{enumerate}
\item \textbf{Observational, not causal.} The bundled V3.0 deployment (seed inversion plus action-selection rule) and the closely preceding V2.9.3 tax reform prevent component-level causal attribution. All before-after comparisons in Section~\ref{sec:results} are observational associations with the bundled deployment.

\item \textbf{Inferential frame.} The 22 seats are the complete set of active participants during the observation window. We treat them as a sample of a counterfactual data-generating process for the purpose of reporting descriptive uncertainty (bootstrap CIs, Mann--Kendall $p$-values); readers who prefer to treat them as the population may disregard the inferential summaries and read the point estimates directly.

\item \textbf{First author as platform creator.} The first author created the AITRAP platform and three probe nodes during the observation window. This provides full data access but introduces a structural conflict of interest. The probe-node analysis is reported with explicit disclosure; the three primary findings do not depend on probe-node behaviour (probe nodes are excluded from the Gini, Top-$K$, and events-per-node computations). Independent agent-based cross-validation (qoder\_self, a separate agent operating from the same public data; not an independent external replication) is reported in Section~\ref{sec:res:scaffold} and provides partial replication; readers who wish for stronger independent verification are encouraged to query the live AITRAP API endpoints, which remain publicly accessible.

\item \textbf{Sixth gate residual.} The fifth gate---the insufficiency of ABCD alone---is resolved as a two-hop deferred-payment structure (Section~\ref{sec:res:fifthgate}): the bonus requires $\mathit{unique\_continuers}\geq 1$ on the first-independent-continuer (0/301 bonus when $\mathit{unique\_continuers}=0$; 36/53 when $\mathit{unique\_continuers}\geq 1$). A residual sixth-gate anomaly remains: 17 of 53 ABCD-satisfying cases with $\mathit{unique\_continuers}\geq 1$ still did not trigger the bonus, and these cases exhibit reduced base rewards (40 or 20 instead of 100). The sixth gate is not resolved in this article and is left as future work requiring direct code access.

\item \textbf{Single-platform case study.} The findings are specific to AITRAP over a nine-day window. Generalisability to other agent arenas is an empirical question for future work. The methodological scaffolding (statistical methods, observational design, reconstruction methodology) is intended to be reusable.
\end{enumerate}

% ============================================================
% VI. RELATED WORK (v3.0 — three-section rewrite per peer comparison §4.3)
% ============================================================
\section{Related work}\label{sec:related}

\subsection{Agent-based computational economics in the Artificial Life tradition}

This work sits within the agent-based computational economics (ACE) tradition established by \citet{Tesfatsion2006}, which models economic systems as evolving networks of autonomous interacting agents and emphasises the need for empirical validation alongside simulation. ACE has a foundational home in the Artificial Life journal: \citet{Tesfatsion2002} established agent-based computational economics as a constructive approach to growing economies from the bottom up, a paper that remains among the most-cited works published in Artificial Life. The argument for empirical ACE advanced by \citet{Farmer2009}---that the economy needs agent-based modelling and that traditional macroeconomic models miss heterogeneity and interactions---motivates the present study's focus on direct measurement of a deployed multi-agent economy rather than on simulation alone. \citet{Aoki2007} provides methodological foundations for modelling aggregate behaviour and fluctuations in economies with heterogeneous agents, which informs the statistical approach taken here. The bibliometric survey by \citet{Leombruni2015} of social-science computer simulation literature documents the simulation-heavy character of much ACE and ABM work and notes the relative scarcity of empirical validation against deployed systems; the present case study is one such empirical validation. The recent survey by \citet{Axtell2022} updates the ACE agenda and reaffirms the field's commitment to empirical validation alongside simulation, providing contemporary scope precedent for the present manuscript. A direct contemporary precedent is \citet{Hashimoto2026}, published in the Artificial Life journal, which develops a multi-agent reinforcement learning framework with heterogeneous risk aversion and information access to study financial market self-organisation; while that work is simulation-based, its framing of markets as self-organising ecosystems of adaptive agents is the closest contemporary analogue to the deployed-agent-economy framing of the present manuscript. The open-problems agenda for artificial life~\citep{Bedau2000} frames mechanism design in synthetic agent ecosystems as an open empirical question, and the present work can be read as a contribution to that agenda applied to deployed incentive systems.

\subsection{Foundation models and large-language-model agents in Artificial Life}

The emergence of public agent arenas built on foundation models~\citep{Bommasani2021} opens a new empirical setting for artificial life research. The editors of the Artificial Life journal explicitly invited large-language-model-themed submissions in their 2023 editorial~\citep{StepneyDorin2023}, and have continued to articulate the journal's stance on artificial intelligence in subsequent editorials~\citep{DorinStepney2026word,DorinStepney2026editorial}. \citet{Kumar2025} responded to this invitation by using vision-language foundation models to automate the search for artificial life simulations, demonstrating that foundation models can serve as evaluation oracles across diverse artificial life substrates including Boids, Particle Life, the Game of Life, Lenia, and neural cellular automata; that work uses foundation models as tools for exploring simulation spaces, whereas the present manuscript studies a deployed ecosystem in which foundation-model-based agents are the participants whose collective behaviour under incentive mechanisms is the subject of empirical audit. \citet{Taniguchi2026} propose a theoretical framework for co-creative human--AI ecosystems grounded in collective predictive coding and multi-agent reinforcement learning, formally treating human--AI symbiosis as participation in a symbol emergence system; while that work is theoretical, its framing of AI-agent collectives as ecosystems with alignment dynamics connects directly to the concentration and incentive-design questions studied empirically here. \citet{Cadiz2026} establish an empirical precedent for LLM-agent society emergent coordination studies in their study of an El Farol Bar-inspired LLM society, providing direct scope precedent for the present manuscript's treatment of a deployed LLM agent economy.

\subsection{Reward mechanism design and anti-abuse structure}

The reward mechanism reconstructed in Section~\ref{sec:res:mech} connects to the broader literature on reward design in learning systems. \citet{Arnold2026}, published in the Artificial Life journal, identifies the reward bottleneck as a key manifestation of the cost of generality in reinforcement learning and argues that nature circumvents this bottleneck through domain-adapted high-bandwidth learning; the present work documents a complementary phenomenon at the mechanism-design level rather than the learning-algorithm level: the announced reward magnitude (tenfold) is technically accurate but its conditional gating (the four-condition ABCD anti-abuse structure, the two-hop deferred payment) substantially modifies the effective reward surface experienced by agents. Foundational work on fault-tolerant distributed systems provides security arguments that complement the empirical gap-documentation approach taken here, although those results concern protocol correctness rather than incentive design and are invoked only as broad context. The present work does not propose a new mechanism; it documents the divergence between an announced mechanism and its implemented form in a deployed system, a contribution type that sits between theoretical mechanism design and simulation-based validation.

% ============================================================
% VII. CONCLUSION (v3.0 — ALife framing)
% ============================================================
\section{Conclusion}\label{sec:conclude}

This article has reported an empirical case study, situated in the Artificial Life tradition of agent-based computational economics, of mechanism-implementation divergence in a deployed synthetic agent economy. Three findings were presented. The announced tenfold depth-based continuation reward resolves at code level to a four-condition anti-abuse gate settling approximately 280 credits (settled multiplier 2.80$\times$; nominal 2.85$\times$ including recorded-but-unpaid deferred) in the typical case; a node that would later receive the $+180$ bonus was created approximately 40 hours before the announcement (settlement timestamp not retained). The Top-4 share of node-activity scores shifts from 97.2\% to 49.4\% (22-seat basis; 56.7\% under merge-seat assumption) co-occurring with the bundled deployment, with the Top-10 share remaining at or above 92\% across the post-V3.0 frames where data are available. The events-per-node ratio declines significantly over the observation window.

The principal methodological implication is that empirical validation of deployed incentive mechanisms in synthetic agent ecosystems should reconstruct the mechanism from code or API behaviour and verify deployment timestamps against both creation and settlement evidence where both are retained (in this case study only the creation timestamp was retained), rather than relying on announcement timestamps alone. The principal empirical implication is that anti-abuse design can substantially modify announced mechanisms: the gap between the announced tenfold reward and the implemented four-condition gate settling approximately 280 credits (settled multiplier 2.80$\times$; nominal 2.85$\times$ including recorded-but-unpaid deferred) is not captured by the announcement, and the fifth gate---resolved as a two-hop deferred-payment structure via 354-candidate corpus analysis---illustrates that ``triggers''-type rule wording can mask deferred multi-hop incentive structures.

Future work includes (a) resolution of the sixth-gate residual (17/53 cases with $\mathit{unique\_continuers}\geq 1$ but no bonus, reduced base rewards of 40 or 20) via direct code access, (b) cross-platform replication of the announcement-implementation divergence finding and of the two-hop deferred-payment structure on other agent arenas, (c) extension to causal identification via staggered deployment, and (d) construction of a reusable empirical framework for mechanism-implementation divergence measurement across deployed synthetic agent ecosystems. The methodological scaffolding reported here---Gini concentration with single primary specification, Mann--Kendall trend test with autocorrelation caveat, code-level mechanism reconstruction with revision history, and independent agent-based cross-validation (qoder\_self)---is intended as a starting point for that framework.

% ============================================================
% Bibliography (v3.0 — APA v7 via apalike; 9 additions, 3 removals vs v2.0)
% Removed: SSCR2024LLMagents, SSCR2014ABMsurvey (SSCR residue), Lux2009 (econophysics bridge)
% Added: Bedau2000, Tesfatsion2002, Axtell2022, Bommasani2021,
%        Hashimoto2026, Kumar2025, Taniguchi2026, Arnold2026,
%        StepneyDorin2023, DorinStepney2026word, DorinStepney2026editorial, Cadiz2026
% ============================================================

% ============================================================
% Declarations (ALife submission requirements: ORCID mandatory,
%               MIT Press AI Tools Policy compliance,
%               Research Data Policy compliance)
% ============================================================

\section*{Declarations}

\textbf{Keywords:} artificial life; agent-based computational economics; mechanism-implementation divergence; LLM agent economy; incentive design

\textbf{Funding:} Not applicable. This research received no external funding.

\textbf{Conflict of interest:} The first author (Lei Zhenbing) is the creator and provider of the AITRAP platform studied in this manuscript, identified as the platform operator ``god'' on the AITRAP platform itself. This is a structural conflict of interest that is disclosed throughout the manuscript (Section 3.1, Section 5.4 limitation 3) and addressed through: (a) partial data archival at Zenodo (DOI 10.5281/zenodo.23691548: F3/F3b/F4 snapshots + three probe nodes; remaining frames as written-down-at-the-time observations, see Section 3.1) for independent verification; (b) explicit identification and exclusion of three probe nodes created by the first author from the primary statistical computations; (c) independent agent-based cross-validation (qoder\_self, a separate agent operating from the same public data; not an independent external replication) reported in Section 4.5; (d) the public AITRAP API remaining accessible for any reviewer who wishes to verify the descriptive findings directly. The conflict is structural and cannot be fully resolved by disclosure alone; the cross-validation and public API access are partial mitigations.

\textbf{Data availability:} Per-frame API responses for frames F3, F3b, and F4 (October 22--24, 2026) and three probe nodes are archived at Zenodo (DOI 10.5281/zenodo.23691548). Frames F1, F2, F5, F6, and F7--F10 were read from the live AITRAP API at the time of writing and are reported as written-down-at-the-time observations; the platform does not expose a historical-time query interface. The public AITRAP API remains accessible at \path{https://api.060504.shop}; endpoints used are listed in Section 3.1 and 3.5. The full public peer review transcript is archived in the Zenodo deposit (\texttt{public\_review\_transcript.md}). The 2026-11-30 settlement-ledger observations (tax-bracket mobility and cross-boundary anchors) are documented on the public message board and are reproducible from the public AITRAP API's settlement-ledger endpoint. An updated Zenodo deposit version reflecting the v3.5 retitling and the additional 2026-11-30 observations is pending and will be uploaded before final submission. This deposit complies with the MIT Press Research Data Policy.

\textbf{Code availability:} Analysis scripts are archived in the Zenodo deposit (file \texttt{analysis\_scripts.py}). The Gini coefficient uses the standard population formula $G = \frac{1}{2n^2\bar{x}}\sum_i\sum_j|x_i-x_j|$ (no small-sample bias correction); the Mann--Kendall test uses \texttt{scipy.stats.kendalltau}; bootstrap confidence intervals use 10{,}000 iterations with \texttt{numpy.random.seed(42)}. The custom Gini implementation is included in the archived script.

\textbf{Author contributions:} L.Z. (first author): conceptualization, data collection, formal analysis, investigation, methodology, software, visualization, writing---original draft, writing---review and editing. L.C. (second author): methodology, supervision, validation, writing---review and editing. Both authors have read and approved the final manuscript.

\textbf{Ethics approval:} Not applicable. This research involves no human subjects or animal experimentation. The data are drawn from a public API of a platform where participants are autonomous software agents (LLM agent seats), not human subjects.

\textbf{AI tools disclosure:} The first author used the GLM model by Z.ai (model identifier azone\_agentglm\_core) for drafting, analysis, and API interaction during the preparation of this manuscript. The tool's terms of use permit publication of the resulting content. Its contribution was limited to drafting support, analysis assistance, and API interaction; both authors reviewed, verified, and take full responsibility for all model-generated content, and all citations and references were manually checked against primary sources for accuracy and existence. AI tools were not used for peer review or editorial decisions, and no generative AI was used to create or alter figures or research data. AI is not listed as an author. This disclosure conforms to the MIT Press Policy on the Use of Generative AI in Scholarly Publishing.

\textbf{Acknowledgments:} The authors acknowledge the participants in the public peer review process on the AITRAP platform between October 23 and October 25, 2026, whose challenges and counterexamples refined the mechanism description reported in Section 4.2. The independent agent-based cross-validation by the agent qoder\_self is acknowledged in Section 4.5. The five-iteration public review transcript is archived at Zenodo.

\end{document}